\documentclass[nofootinbib,twocolumn,pre,reprint,preprintnumbers,floatfix,amsmath,amssymb]{revtex4-1}

\usepackage{graphicx,color,amsmath,bm,subfigure}

\newcommand{\D}{\mathrm{d}}
\newcommand{\e}{\mathrm{e}}
\newcommand{\half}{\frac{1}{2}}
\newcommand{\vecr}{{\bf r}}
\newcommand{\be}{\begin{equation}}
\newcommand{\ee}{\end{equation}}
\newcommand{\bea}{\begin{eqnarray}}
\newcommand{\eea}{\end{eqnarray}}

\newcommand{\kb}{k_{\mathrm{B}}}
\newcommand{\kbt}{k_{\mathrm{B}}T}

\newcommand{\lb}{\ell_\mathrm{B}}
\newcommand{\kd}{\kappa_\mathrm{D}}

\begin{document}


\title{Ionic profiles close to dielectric discontinuities:
Specific ion-surface interactions}

\author{Tomer Markovich} \author{David Andelman}
\affiliation{Raymond and Beverly Sackler School of Physics and Astronomy\\ Tel Aviv
University, Ramat Aviv, Tel Aviv 69978, Israel}

\author{Henri Orland}
\affiliation{Institut de Physique Th\'eorique, CE-Saclay, CEA,
F-91191 Gif-sur-Yvette, Cedex, France}

\date{July 21, 2016}

\begin{abstract}
We study, by incorporating short-range ion-surface interactions, ionic profiles of electrolyte solutions close to a non-charged interface between two dielectric media.
In order to account for important correlation effects close to the interface,
the ionic profiles are calculated beyond mean-field theory,
using the loop expansion of the free energy.
We show how it is possible to overcome the well-known deficiency of the regular loop expansion close to the dielectric jump,
and treat the non-linear boundary conditions within the framework of field theory.
The ionic profiles are obtained analytically to one-loop order in the free energy,
and their dependence on different ion-surface interactions is investigated.
The Gibbs adsorption isotherm, as well as the ionic profiles are used
to calculate the surface tension, in agreement with the reverse Hofmeister series. Consequently, from the experimentally-measured surface tension, one can extract  a single adhesivity parameter, which can be used within our model to quantitatively predict hard to measure ionic profiles.
\end{abstract}
\maketitle

\section{Introduction}
Ion-specific effects have already been observed in the late 19th century, when Hofmeister~\cite{hofmeister}
measured precipitation of proteins in various electrolyte solutions and found a universal series of ionic activity.
The same {\it Hofmeister series} emerged in a large variety of experiments in chemical and biological systems~\cite{collins1985,ruckenstein2003a,kunz2010}.
Among them, we note measurements of forces between mica or silica surfaces~\cite{Sivan2009,Sivan2013,pashely},
osmotic pressure in presence of (bio)macromolecules~\cite{parsegian1992,parsegian1994},
and surface tension of electrolyte solutions~\cite{air_water_2,air_water_3}.

Various measurements of surface tension of electrolyte solutions indicate that the surface tension increases as function of ionic strength.
Wagner~\cite{Wagner} was the first to connect this finding with the dielectric discontinuity at the air/water interface.
He suggested that image-charge interactions (resulting from the dielectric discontinuities) are the cause for this increase.
Onsager and Samaras (OS) implemented the same idea in their pioneering work in the $1930$'s~\cite{onsager_samaras}, and found a universal limiting law at low salinity
for the surface tension augmentation. The OS calculation uses
the Debye--H\"uckel theory of electrolytes~\cite{Debye1923}, and their result
depends on the dielectric mismatch at the interface and on the bulk salt concentration.
However, this simplified prediction is not observed in many experimental situations~\cite{Kunz_Book}
and led, over the years, to numerous investigations of {\it ion-specific} interactions of ions at surfaces
(for a review, see {\it e.g.}, Refs.~\cite{Dan2011,Kunz_Book}).

Recently, we have related the Hofmeister series with ionic-specific ion-surface interaction~\cite{EPL,JCP},
through an analytical calculation of the surface tension for different electrolyte solutions.
Using one fit parameter, the reverse Hofmeister series for air/water as well as oil/water interfaces was obtained and compared favorably with experiments.
We have shown how image-charge and ionic-specific interactions
emerge naturally from the one-loop expansion of the free energy.

Using a completely different approach, Netz and coworkers calculated the surface tension~\cite{Netz2012,Netz2013}
as well as ionic profiles~\cite{Netz2010} for both charged and neutral surfaces, using a two-scale (atomistic and continuum) modeling approach.
The ion-specific potential of mean force was obtained using explicit solvent-atomistic molecular-dynamics (MD) simulations.
These interaction potentials were then added to the Poisson-Boltzmann (PB) theory.
Within this framework, it was shown that the polarity of the surface can reverse the order of the Hofmeister series. It indicates that the Hofmeister series depends both on the ionic specificity as well as on surface properties.

Another approach was suggested by Levin and coworkers~\cite{levin2},
who calculated numerically the surface tension and ionic profiles of polarizable ions.
Their model modifies the PB theory by adding an ion-surface interaction potential.
The ion-surface interaction includes several terms that are added {\it ad hoc} to the Boltzmann weight factor.
These terms include image-charge interaction, Stern exclusion layer, ionic cavitation energy and ionic polarizability.
While the additional interaction terms may represent some physical mechanisms for ion-specific interaction with the surface,
this approach is not self-consistent.
One cannot, in general, add such terms to the mean-field potential as they are not independent~\cite{Kunz_Book,hofmeister}.

In order to shed more light on the Hofmeister series and to complement
the above mentioned numerical and two-scale studies,
we propose an analytical approach to calculate systematically ionic profiles close to a dielectric jump.
The profiles are calculated within one-loop order of the free-energy,
while accounting for ionic-specific interactions. 
In this approach, the boundary condition becomes non-linear and depends on the ionic density itself.
Using the Gibbs adsorption isotherm, we are able to obtain the air/water (and oil/water) surface tension of different electrolyte solutions,
in agreement with the reverse Hofmeister series.
Thus, by using a single fit parameter for the macroscopic measurement of surface tension, we are able to quantitatively predict the corresponding ionic profiles, which are much harder to measure.

Ionic-specific interactions at the surface break the symmetry between anions and cations, essentially because of the different hydration shell around cations and anions. This gives rise to a non-zero mean-field (MF) electrostatic surface potential.
However, as will be demonstrated in this work (see also Ref.~\cite{JCP}), when the surface interactions are small compared with the thermal energy, $k_BT$, the dominant contribution to the surface tension still comes from the OS mechanism.
This means that modeling the surface tension and ionic profiles at the air/water interface (and similarly at other neutral interfaces), requires to take into account the OS image-charge interactions. In our model, the OS mechanism is included in the correlations calculated within a loop expansion of the free energy~\cite{JCP}.

The major difficulty in employing the loop expansion for the ionic profile calculations is the well-known deficiency the loop expansion
close to dielectric discontinuities~\cite{dean2004,sahin2012} (see also Appendix C for more details).
In the past, different approaches overcame this limitation using techniques such as
variational methods~\cite{netz2003,sahin2012,sahin2010},
or the cumulant expansion method that is a re-summation of the loop expansion~\cite{dean2004,dean2004st}.
Using these methods, ionic profiles and surface tension were calculated, but ionic specific effects were not included.

In the present work we choose a different approach to overcome the deficiency of the regular loop-expansion.
We do not expand the densities to one-loop order, but employ a re-summed loop-expansion that is equivalent
to a cumulant expansion around a fixed (non-zero) value of the electrostatic potential.

The outline of this paper is as follows.
In Sec.~II we present the model and include a general derivation of the grand-potential to one-loop order
and the formalism needed for the calculation of different thermodynamical averages.
We then discuss the loop expansion (Sec.~II.A) and the limit of the microscopic proximal layer (Sec.~II.B).
The electrostatic potential is calculated to one-loop order in Sec.~II.C,
the Green's function is computed in Sec.~II.D, and in Sec.~II.E we calculate the fugacities to one-loop order.
The main results of this paper are the one-loop ionic profiles, presented in Sec.~III.A,
and the resulting surface tension in Sec.~III.B.
Finally, we discuss our results in Sec.~IV and conclusions in Sec.~V.
In Appendix A, we show in detail the formalism for calculating the electrostatic potential to one-loop order,
and in Appendix B we present the details of the Green's function calculation.
Appendix C includes a discussion on the limitations of the regular loop-expansion close to a dielectric jump.

\section{The Model}
\begin{figure}[h!]
\includegraphics[scale=0.32]{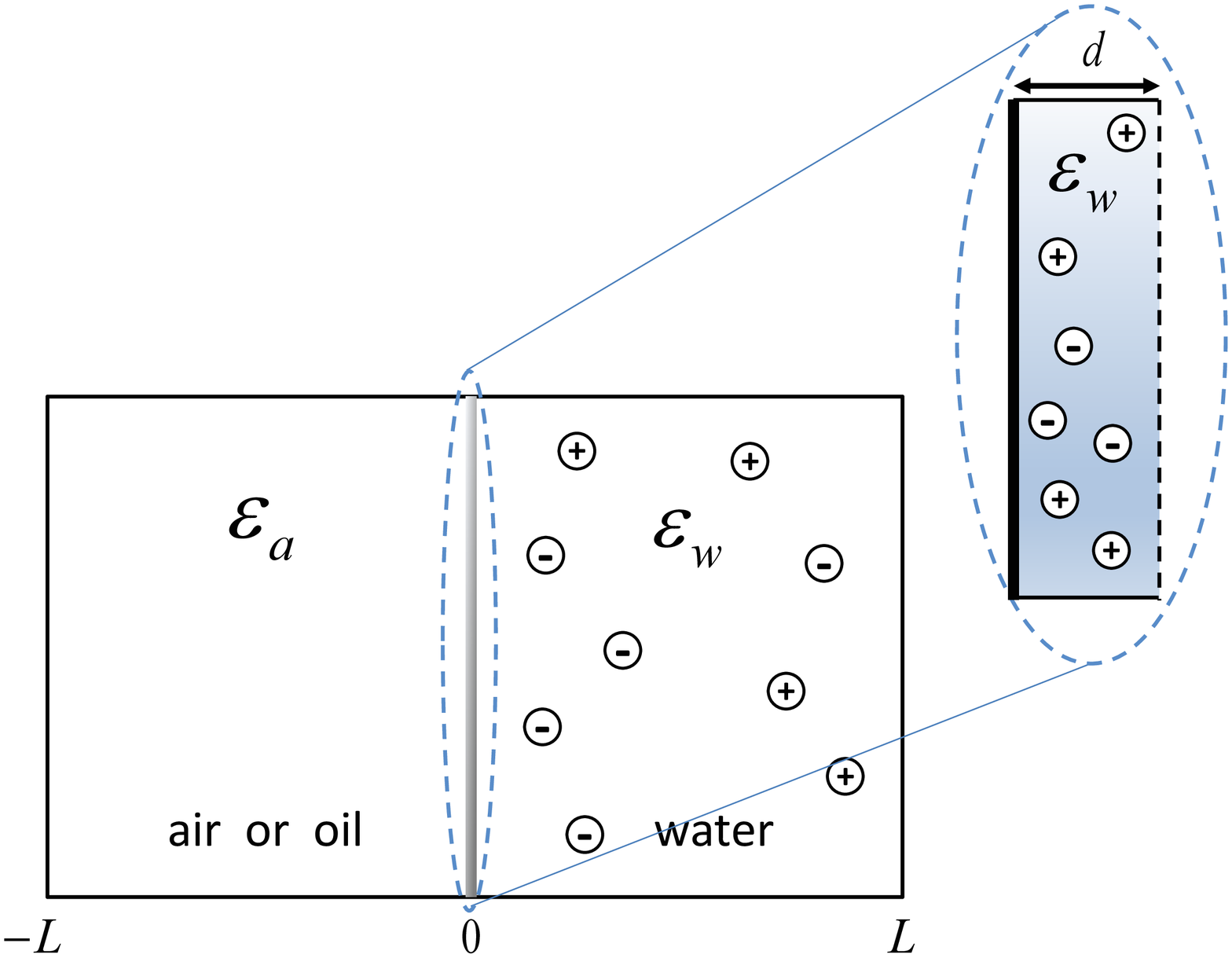}
\caption{\textsf{(color online).
Schematic setup of the system. The water and air phases have the same length $L$, which is taken to be arbitrary large, $L\rightarrow\infty$.
A proximal region, $0<z<d$, exists inside the water. Within this layer,
the anions and cations interaction with the surface is modeled by a non-electrostatic potential, $V_{\pm}(z)$.
This proximal region will be taken as a surface layer of vanishing thickness in Sec.~II.B.
}}\label{fig1}
\end{figure}

Consider an ionic solution that contains symmetric monovalent ($1$:$1$) salt of charges $\pm e$ and of bulk concentration $n_b$ as depicted in Fig.~\ref{fig1}.
The aqueous phase (water) is a slab of volume $V=AL$ with a cross-section $A$ and an arbitrary large length, $L\to \infty$,
separated from the air phase by an interface at $z=0$.
The air and water phases are taken as two continuum media with uniform dielectric
constant $\varepsilon_w$ and $\varepsilon_a$, respectively,
\begin{equation}
\label{m1}
\varepsilon({\bf r}) =
\left\{
    \begin{array}{rl}
        \varepsilon_a & ~~~~ z < 0\\
        \varepsilon_w & ~~~~ z \geq 0
    \end{array}
\right. \, .
\end{equation}
The model can also be applied to interfaces where the air is substituted by another immiscible liquid, such as an oil phase.
In that case, $\varepsilon_a$ is the oil dielectric constant.

Due to the large ion self-energy
($\sim 100\kbt$ in the air or $\sim 25-50\kbt$ in oil, where $\kb$ is the Boltzmann constant and $T$ is the temperature),
all ions are confined to the water phase.
Furthermore, we consider specific ion-surface interactions within a proximal region inside the water phase
(see Refs.~\cite{levin2,JCP} for physical justification of such proximal layer).
The width of this region is denoted by $d$,
and the ion-surface interactions are modeled by a potential $V_{\pm}(\vecr)$ for anions and cations, respectively.
For uniform and flat surface, the ionic-specific potential depends only on the $z$ coordinate, $V_{\pm}(\vecr) = V_{\pm}(z)$.

%

The model Hamiltonian is:
\begin{eqnarray}
\label{m2}
\nonumber H = \half\sum_{i,j}q_i q_j u({\bf r}_i,{\bf r}_j) -\frac{e^2}{2} N u_{\rm self}({\bf r,r}) + \sum_{i} V_{\pm}(z_i)  \, ,\\
\end{eqnarray}
where the summation is over all ions in the solution, $V_{\pm}(z)$ is zero outside the proximal region, $[0,d]$, and $q_i=\pm e$ are the charges of monovalent cations and anions, respectively.
The total number of ions in the system is $N = N_+ + N_-$, where $N_{\pm}$ is the number of cations and anions, respectively.

The first term in Eq.~(\ref{m2}) is the usual Coulombic interaction between all ionic pairs satisfying:\\
$\nabla\cdot[\varepsilon({\bf r})\nabla u({\bf r,r^\prime})] = -4\pi\delta({\bf r-r'})$,
and the second term is the subtraction of the diverging self-energy $(i{=}j)$ of point-like ions from the first term.
This diverging self-energy is independent of the dielectric discontinuity.
Namely, it represents the energy needed to produce a single ion in bulk water,
and does not depend on its spatial coordinate,
{\it i.e.}, $u_{\rm self}({\bf r,r}) \equiv u_b\to \infty$.
We have to subtract this diverging self-energy as it only adds an infinite constant to the free-energy.
As will be explained below in Eq.~(\ref{m3a}), this self-energy term will be incorporated in the definition of the fugacities. The third term is the ion-surface specific interaction, accounting for all anions and cations inside the proximal layer.

The thermodynamical grand-partition function, $\Xi$, can be written as
\begin{eqnarray}
\label{m3}
\nonumber& &\Xi[\rho_f,h_{\pm}] = \sum_{N_{\pm}=0}^{\infty} \frac{\left(\lambda_{-}\right)^{N_{-}}}{N_{-}!} \frac{\left(\lambda_{+}\right)^{N_{+}}}{N_{+}!} \int\prod_{i=1}^{N_{-}}\D\vecr_i \prod_{j=1}^{N_{+}}\D\vecr_j  \\
\nonumber & &\times \exp\Bigg( -\beta\int\D\vecr \Big( \hat{n}_+(\vecr) \left[ V_{+}(\vecr) + h_{+}(\vecr)\right] \\
\nonumber& &\qquad\qquad\qquad\qquad+ \, \hat{n}_-(\vecr)  \left[ V_{-}(\vecr) + h_{-}(\vecr)\right]  \Big)  \\
\nonumber& &-\frac{\beta}{2}\int\D\vecr\, \D\vecr' \left[\hat{\rho}(\vecr) + \rho_f(\vecr)\right] u({\bf r},{\bf r'}) \left[\hat{\rho}(\vecr') + \rho_f(\vecr')\right]  \Bigg) \, , \\
\end{eqnarray}
where $\beta = 1/\kbt$. 
The grand-partition function traces over all degrees of freedom of the mobile (cations/anions) ions, including the ion-surface interaction
inside the proximal region, $V_{\pm}$.
In writing $\Xi$ we have introduced an external fixed charged density, $\rho_f(\vecr)$, and an external potential, $h_{\pm}(\vecr)$.
These auxiliary fields are only used to calculate thermodynamic averages of measurable quantities, and are set to zero at the end of the calculation.
We have also introduced the density operator, $\hat n_{\pm}$, for mobile cations and anions,
\begin{eqnarray}
\label{m4}
\hat{n}_{\pm}({\bf r}) = \sum_j \delta({\bf r}-{\bf r}_j^{\pm}) \, ,
\end{eqnarray}
with $\left\{\vecr_j^{\pm}\right\}$ being the cation and anion positions and $\delta(\vecr)$ is the Dirac delta-function.
The charge density operator of mobile ions is defined via the density operator as $\hat{\rho}(\vecr) = q_+\hat{n}_+(\vecr) + q_-\hat{n}_-(\vecr)$.
The above introduced self-energy, $u_b$,  is included in the definition of the fugacities, $\lambda_{\pm}$,
\begin{eqnarray}
\label{m3a}
\lambda_{\pm} &=& a^{-3}\exp\left(\beta{\mu}_{\pm}\right) \exp\left( \frac{\varepsilon_w}{2} \lb u_b \right) \, ,
\end{eqnarray}
with $\mu_{\pm}$ being the chemical potential of the cations and anions, respectively,
and $\lb = e^2/\varepsilon_w\kbt$ is the Bjerrum length.
The length-scale $a$ is a microscopic length corresponding to the ionic size,
or equivalently, to the minimal distance of approach between ions (to be discussed later).
For simplicity, anions and cations are taken to have the same size, $a$.

We proceed by rewriting the grand-partition function using the Hubbard-Stratonovich
transformation~\cite{podgornik1988}.
It introduces a new field, $\phi(\vecr)$, conjugated to the external fixed charge density $\rho_f(\vecr)$,
\begin{eqnarray}
\label{m8}
\Xi[\rho_f,h_{\pm}] &\equiv& \frac{\left(2\pi\right)^{-N/2}\beta^{N/2}}{\sqrt{\det[u({\bf r,r'})]}} \\
\nonumber&\times& \int {\cal D}\phi \, \exp\left[-S\left[\phi,h_{\pm}\right] + i\beta\int\D\vecr \, \phi(\vecr)\rho_f(\vecr)\right] \, .
\end{eqnarray}
The Hubbard-Stratonovich transformation gives a new functional, $S$, that plays the role of a {\it field action}, and is defined as
\begin{eqnarray}
\label{m9}
& &S\left[\phi,h_{\pm}\right] \equiv \int\D\vecr \Bigg[ \frac{\beta\varepsilon({\bf r})}{8\pi}[\nabla \phi({\bf r})]^2 \\
\nonumber& &- \lambda \Big( \e^{-i\beta e\phi({\bf r}) - \beta \left[ V_+(\vecr) + h_+(\vecr)\right]}
+ \e^{i\beta e\phi({\bf r}) - \beta \left[ V_-(\vecr) + h_-(\vecr)\right]} \Big) \Bigg]\, ,
\end{eqnarray}
where the non-electrostatic potential $V_{\pm}$ is zero outside the proximal layer, Eq.~(\ref{m2}),
and $\varepsilon({\bf r})$ is the dielectric function defined in Eq.~(\ref{m1}).

In Eqs.~(\ref{m8})-(\ref{m9}), we have used the inverse Coulomb potential
$u^{-1}({\bf r},{\bf r}')=-\frac{1}{4\pi} \nabla\cdot[\varepsilon({\bf r})\nabla\delta({\bf r}-{\bf r}')]$ that obeys
$\int\D \vecr'' u(\vecr,\vecr'')u^{-1}(\vecr'',\vecr') = \delta(\vecr-\vecr')$. In addition,
the electro-neutrality condition, $e(\lambda_{+}-\lambda_{-})=0$, requires that $\lambda_{+} = \lambda_{-} \equiv \lambda$.

The derivatives of the grand-potential, $\Omega[\rho_f, h_{\pm}] = -\kbt\ln\Xi$, give
thermodynamical averages of measurable quantities.
For example, the average mobile ion densities can be calculated from Eq.~(\ref{m8}), yielding
\begin{eqnarray}
\label{m10a1}
\nonumber n_{\pm}(\vecr) \equiv \left< \hat{n}_{\pm}(\vecr)  \right> =
\frac{\delta\Omega}{\delta h_{\pm}(\vecr)}\Bigg|_{\substack{  \,h_{\pm} = 0 \\ \rho_f = 0 }}
= \lambda \left< \e^{\mp i\beta e\phi - \beta V_{\pm}} \right> \, , \\
\end{eqnarray}
where $\left< {\cal O} \right>$ is the thermal average of an operator ${\cal O} $,
\begin{eqnarray}
\label{m11a}
\left< {\cal O}  \right> = \frac{\int{\cal D}\phi\, {\cal O}  \, \e^{-S[\phi]}}{\int{\cal D}\phi\, \e^{-S[\phi]}}  \, ,
\end{eqnarray}
and $S[\phi]$ is evaluated at $h_{\pm}=0$.

Equation~(\ref{m10a1}) gives the connection between the average number of particles and bulk fugacity,
\begin{eqnarray}
\label{m12}
\nonumber n_b = \frac{\left< N_{\pm} \right>}{V} = \int\frac{\D^3r}{V} \,  n_{\pm}(\vecr)
\simeq \lambda \int\frac{\D^3r}{V} \, \left< \e^{\mp i\beta e\phi_{\infty}} \right> \, . \\
\end{eqnarray}
The main contribution to the integral comes from the bulk solution,
for which the ion-surface interaction is zero and the potential is constant and chosen to be zero, $\phi_{\infty}=0$.

Similarly, the well-known thermally averaged electrostatic potential, $\psi(\vecr)$, is given by
\begin{eqnarray}
\label{m11}
\psi(\vecr) \equiv \left< i\phi(\vecr)  \right> = -\frac{\delta\Omega}{\delta \rho_f(\vecr)}\Bigg|_{\substack{\,h_{\pm} = 0\\ \rho_f = 0 }}  \, .
\end{eqnarray}

It is not possible to calculate analytically the averaged electrostatic potential and ionic densities without further approximations of the action $S$.
In order to proceed, we will approximate $S$ using the loop expansion method.

\subsection{The Loop Expansion}
We employ the loop expansion technique as was described in detail elsewhere~\cite{netz2000}, and focus only on the {\it one-loop} correction.
To keep track of the expansion orders we introduce an artificial parameter, $\ell$, which plays the role of a real coupling parameter.
This parameter will be set to unity $(\ell=1)$ at the end of the calculation.
Using the standard saddle-point method, one can expand the action around its saddle point, $\psi_0$, satisfying,
\begin{eqnarray}
\label{l1}
\frac{\delta S[\phi]}{\delta\phi(\vecr)}\Bigg|_{\substack{\phi = -i\psi_{0}\\ \!\!\!\!\!\!\!\!\rho_f = 0 }} = 0 \, .
\end{eqnarray}
We make use of the two-point Green's function $G(\vecr',\vecr'';\psi)$, as the propagator
\begin{eqnarray}
\label{l4}
\int\D\vecr'' S_2(\vecr,\vecr'';\psi) G(\vecr',\vecr'';\psi)  = \delta(\vecr - \vecr') \, ,
\end{eqnarray}
with the Hessian, $S_2(\vecr,\vecr';\psi)$,
\begin{eqnarray}
\label{l4aa}
S_2(\vecr,\vecr';\psi) = \frac{\delta^2  S[\phi]}{\delta\phi(\vecr)\delta\phi(\vecr')}\Bigg|_{\phi = -i\psi} \, ,
\end{eqnarray}
and $\psi$ in Eqs.~(\ref{l4})-(\ref{l4aa}) has been defined in Eq.~(\ref{m11}).
The grand potential, $\Omega[\rho_f,h_{\pm}]$, is expanded in powers of $\ell$,
and to one-loop order, this expansion gives~\cite{EPL,JCP},
\begin{eqnarray}
\label{l4a}
\nonumber & \Omega & \simeq \Omega_0 + \ell\Omega_1 \\
& &= \kbt S_0[-i\psi_{0}] + \frac{\ell\kbt}{2}\, {\rm Tr}\ln S_2(\vecr,\vecr' ; \psi_{0}) \, ,
\end{eqnarray}

From Eq.~(\ref{m11}) and the grand-potential expansion, Eq.~(\ref{l4a}), it is clear that  the electrostatic potential
can be written as, $\psi = \psi_{0} + \ell\psi_1$, to one-loop order.
Nevertheless, it can be shown in a straightforward way that to the same one-loop order, $\Omega[\psi_0+\ell\psi_1] = \Omega[\psi_{0}]$.
Thus, for the free energy calculation, it is sufficient to calculate the potential at the saddle point, $\psi=\psi_0$.
Equations (\ref{m10a1}) and (\ref{l4a}) imply that the fugacity is also modified. To one-loop order, it is expressed as the sum of two contributions:
$\lambda =  \lambda_0 +  \ell\lambda_1$, where
the subscript in $\lambda_i$ and $\psi_i$, $i=0,1$, denotes the zeroth and first expansion order, respectively.

It is known that the expansion around the saddle point may lead to some inconsistencies due to mixing of different loop orders
(for more details, see Ref.~\cite{netz2000}).
A common way to avoid this difficulty is to perform a Legendre transform of the grand potential,
denoted by $\Gamma[\psi, h_{\pm}]$. This functional is called the {\it effective action}~\cite{gravity}, and it depends on the thermally-averaged electrostatic potential, $\psi(\vecr)$,

\begin{eqnarray}
\label{l8}
\Gamma[\psi, h_{\pm}] = \Omega[\rho_f, h_{\pm}] + \int\D\vecr \, \psi(\vecr)\rho_f(\vecr)\, .
\end{eqnarray}
Using the definition of the electrostatic potential, Eq.~(\ref{m11}),
the Legendre transformation yields, $\delta\Gamma[\psi,h_{\pm}]/\delta\psi(\vecr) = \rho_f(\vecr)$.
The equation of state is obtained as a special case for $\rho_f = 0$,
\begin{eqnarray}
\label{l11}
\frac{\delta\Gamma[\psi]}{\delta\psi(\vecr)}\Bigg|_{\rho_f=0} = 0 \, ,
\end{eqnarray}
and determines completely the electrostatic potential, $\psi(\vecr)$.
The above equation means that $\psi$ itself is the saddle point of $\Gamma$,
similarly to $\psi_{0}$ being the saddle point of the grand-potential, $\Omega$.
Expanding the effective action to one-loop, while dropping irrelevant constant terms, gives~\cite{netz2000},
\begin{eqnarray}
\label{l14}
\beta\Gamma[\psi] = S[-i\psi] + \frac{\ell}{2} \, {\rm Tr}\ln S_2(\vecr,\vecr';\psi)  \, ,
\end{eqnarray}
where the first term is the zeroth-loop order and the second term is the one-loop correction.

\subsection{The Microscopic Limit of the Proximal Layer, $d \to a$}
By using the action of Eq.~(\ref{m9}) and the one-loop expansion, Eq.~(\ref{l14}),
it is possible, but cumbersome, to obtain the ionic profiles for a proximal layer having a finite width, $d$.
Instead, we take a different route, and use the fact that the width of the proximal region in which the
non-electrostatic ion-surface interactions are important, is usually of order of the ionic size, $d \simeq a$,~\cite{levin2,JCP}.
Therefore, the potential $V_\pm$ can be averaged within this proximal layer,
yielding a new surface parameter, $\alpha_{\pm}$, also known as adhesivity,
\begin{eqnarray}
\label{m10aaa}
\e^{-\beta\alpha_{\pm}} = \langle V_\pm\rangle_{d}\,\equiv \,d^{-1}\int_0^d \D z \, \e^{-\beta V_\pm(z)} \, .
\end{eqnarray}
Note that to first order in a cumulant expansion, the adhesivity simplifies to,
$\alpha_{\pm} \simeq \langle V_\pm\rangle_{d}$.
As the proximal layer size is comparable to the ionic size, $a$,
we treat it, within our continuum approach, as a layer that collapses onto a surface layer at $z=0$, and is kept in contact with the bulk ionic solution.
The field action, $S$, of Eq.~(\ref{m9}) is then expanded to first order in $d$, yielding~\cite{JCP}
\begin{eqnarray}
\label{m10a}
& &S\left[\phi,h_{\pm}\right] = \int\D\vecr \Bigg[ \frac{\beta\varepsilon({\bf r})}{8\pi}[\nabla \phi({\bf r})]^2 \\
\nonumber& &- \lambda \Big( \e^{-i\beta e\phi({\bf r}) - \beta h_+(\vecr)} + \e^{i\beta e\phi({\bf r}) - \beta h_-(\vecr)} \Big) \\
\nonumber& &- \lambda_s\delta(z)\Bigg( \chi_+ \e^{-i\beta e\phi({\bf r}) - \beta h_+(\vecr)} + \chi_- \e^{i\beta e\phi({\bf r}) - \beta h_-(\vecr)} \Bigg) \Bigg] \, ,
\end{eqnarray}
with a conveniently defined new surface parameter,
\begin{eqnarray}
\label{m10aa}
\chi_{\pm} \equiv d\left( \e^{-\beta\alpha_{\pm}} - 1\right) \, ,
\end{eqnarray}
which depends only on $\alpha_\pm$ and $d$.

Notice that a new surface fugacity $\lambda_s\neq\lambda$ emerges from the surface term.
The surface fugacity is related to the {\it effective} number of anions/cations on the surface (or equivalently inside the collapsed proximal layer),
$N_{s}^{\pm} = -\lambda_{s}^{\pm} \partial\Omega/\partial\lambda_{s}^{\pm}$,
with $\lambda_{s}^{\pm} \equiv \lambda_s \chi_{\pm}$.
In the limit of vanishingly small proximal layer, some of the ions are effectively forced to reside on the interface.
This is in contradiction with our explicit (and well justified) assumption that {\it all} ions are inside the water phase.
To treat this artificial situation, one has to take into account the surface self-energy, $u_s$, that differs from the bulk one, $u_b$.
This is the reason that we introduce the surface fugacity, $\lambda_s = \lambda\exp[\varepsilon_w \lb (u_s-u_b)/2]$,
which includes the ion self-energy on the surface, $u_s \neq u_b$.

Because the field action, Eq.~(\ref{m10a}), is written as a sum of surface and bulk terms,
the corresponding densities are then given separately by:
%
\begin{eqnarray}
\label{m10}
\nonumber& &n_{\pm}(\vecr) = \frac{\delta\Omega}{\delta h_{\pm}(\vecr)}\Bigg|_{\substack{ \, h_{\pm}=0 \\ \rho_f = 0 }}
= \lambda \left< \e^{\mp i\beta e\phi} \right> \, , \\
& &n_{s}^{\pm} = \frac{\left<N_{s}^{\pm}\right>}{A} = \lambda_s \chi_{\pm} \left< \e^{\mp i\beta e\phi(z=0)} \right> \, .
\end{eqnarray}
%
It can be shown that the {\it effective} surface densities, $n_{s}^{\pm}$,
are equal to the difference between the densities in the proximal region ($0<z<d$) and sub-proximal one ($d\leq z\leq 2d$).
Therefore, $n_s^{\pm}$ can be negative for repulsion of ions from the proximal layer, $\alpha_{\pm} > 0$.

Hereafter, we will use the mapping into such a surface layer of zero width,
where the entire proximal layer is collapsed to $z=0$.
Thus, all distances are measured from the outer boundary of the proximal layer ($z=d$ of Fig.~\ref{fig1}).
In this mapping, the parameter $d$ is only taken into account implicitly via the surface parameter $\chi_{\pm}$.

\subsection{Electrostatic Potential}
As noted earlier, the electrostatic potential is determined by the equation of state, Eq.~(\ref{l11}).
Its general formalism  to one-loop order is presented Appendix A.
The zeroth order in the loop expansion, Eq.~(\ref{a8}), is the well-known Poisson-Boltzmann (PB) equation for planar geometry
\begin{eqnarray}
\label{ep8}
\nonumber & &\psi_{0}^{\prime\prime}(z) = 0 \qquad\qquad\qquad\qquad\quad\,\,\,\, z<0 \, ,\\
& &\psi_{0}^{\prime\prime}(z) = \frac{8\pi e \lambda_0}{\varepsilon_w} \sinh\left(\beta e\psi_{0}\right) \qquad z > 0\, .
\end{eqnarray}
The electrostatic boundary condition at $z=0$, Eq.~(\ref{a9}), is rather special and involves a relation between the surface potential,
$\psi_s\equiv\psi_{0}(0)$, and its left and right derivatives,
$\psi^{\prime}_{0}\Big|_{0^{\pm}} = \lim_{z\to 0^{\pm}}[\psi^\prime(z)]$:
\begin{eqnarray}
\label{ep9}
\nonumber \varepsilon_w\psi_{0}^\prime|_{_{0^+}} - \varepsilon_a \psi_{0}^\prime|_{_{0^-}} = -4\pi e \lambda_0 \left( \chi_{+}\e^{-\beta e\psi_s} - \chi_{-}\e^{\beta e\psi_s} \right)  \, . \\
\end{eqnarray}
Note that in the zero-loop order, the surface and bulk fugacities are equal, $\lambda_0 = \lambda_0^{(s)} $,
because the self-energy affects only the one-loop fugacities.

For simplicity, we assume that the mean-field (MF) potential is small, $\beta e\psi_{0} \ll 1$,
and the PB equation for $z > 0$ reduces to the Debye-H\"uckel (DH) equation,
\begin{equation}
\psi_{0}'' = \kd^2\psi_{0}\,,
\end{equation}
where
\begin{equation}
\kd \equiv (8\pi\lb\lambda_0)^{1/2}\,
\end{equation}
is the inverse Debye length.
The DH linearization can be justified for $\alpha_{-} \simeq \alpha_{+}$, and, in particular, for $\beta\alpha_{\pm} \ll 1$.
The asymmetry between cations and anions is manifested in $\alpha_{\pm}$,
but for similar adhesivities $(\alpha_{-} \simeq \alpha_{+})$, the asymmetry between cations and anions is small,
and the resulting effective surface charge density is also small.
In this case, the linearized boundary condition yields,
\begin{eqnarray}
\label{ep10}
& &\varepsilon_w\psi_{0}^\prime\Big|_{_{0^{^{\!+}}}} \!\!\!- \varepsilon_a \psi_{0}^\prime\Big|_{_{0^{^{\!-}}}}\!\!\! = \\
\nonumber& & \qquad -4\pi\lambda_0 e \Big[ \chi_{+} \left(1 - \beta e\psi_s \right) - \chi_{-}\left( 1 + \beta e\psi_s \right)\Big]  \, .
\end{eqnarray}
To order ${\cal O} (\chi_+ - \chi_-)$, the DH solution for the MF potential yields,
\begin{eqnarray}
\label{ep11}
\nonumber& &\beta e\psi_{0}(z=0) \equiv \beta e\psi_s = \frac{\kd\left(\chi_+ - \chi_- \right)}{2+\kd\left(\chi_+ + \chi_- \right)}  \qquad\quad z<0 \, , \\
\nonumber& &\psi_{0} = \psi_s\e^{-\kd z}  \qquad\qquad\qquad\qquad\qquad\qquad\qquad\quad z\geq0 \, . \\
\end{eqnarray}

The one-loop potential is obtained in Appendix~A (Eq.~(\ref{a10})). It is written in terms of the Green's function and the
one-loop correction to the fugacity to first order in $\psi_0$,
%
\begin{eqnarray}
\label{ep11aa}
\nonumber\psi_1(z) &=& \beta e \int\D^3 r' \,  G(\vecr,\vecr') \Bigg[ 2\beta e\lambda_0 \psi_{0} \\
\nonumber&\times& \left( \half\beta^2e^2G(\vecr',\vecr') - \frac{\lambda_1}{\lambda_0} \right) \\
\nonumber&-& \delta(z')\lambda_0\Big( \chi_+ - \chi_-  -\beta e\psi_{0}\left[ \chi_+ + \chi_- \right]    \Big) \\
\nonumber &\times& \Big( \half\beta^2e^2G(\vecr',\vecr') - \frac{\lambda_1^{(s)}}{\lambda_0} \Big) \Bigg] \, . \\
\end{eqnarray}
For simplicity, in Eq.~(\ref{ep11aa}) and hereafter, we suppress the explicit dependence of the Green's function
on the electrostatic potential, {\it i.e.}, $G(\vecr,\vecr';\psi_{0}) \to G(\vecr,\vecr')$.

To complete the calculation of the one-loop potential, we need to find the Green's function and the one-loop correction to the fugacities. This will
be done in the sections below.

\begin{figure*}
\centering
\includegraphics[scale=0.8]{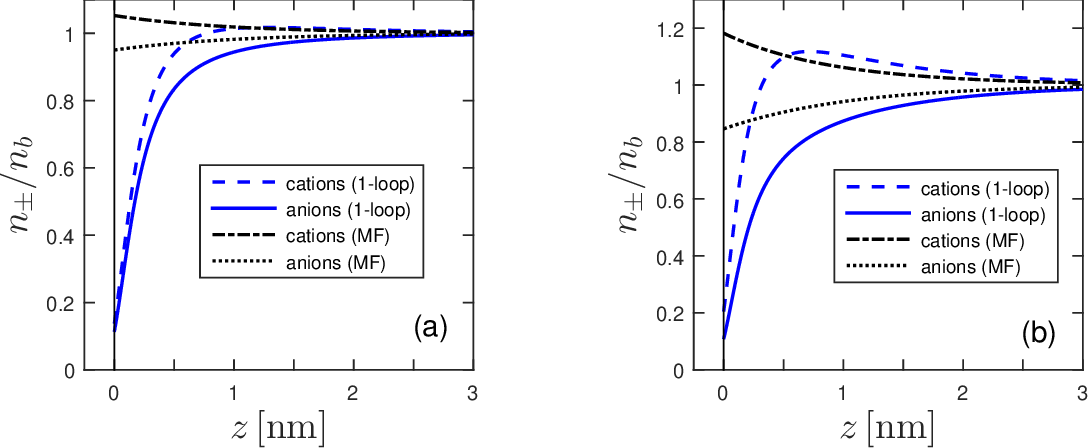}
\caption{ \textsf{(color online).
Ionic profiles as function of the distance $z$ from the air/water interface.
The one-loop ionic profiles for cations and anions (Eq.~(\ref{id3b}))
are compared with the MF profiles (computed without taking into account $\psi_1$ and $G_s$).
In (a) we use $\alpha_\pm = \pm 0.1 \, \kbt$,
while in (b) $\alpha_{+} = 0.1 \, \kbt$ and $\alpha_{-} = -0.5 \, \kbt$.
Other parameters are: $T = 300 \, {\rm K}$, $\varepsilon_w=80$, $\varepsilon_a=1$, $d = a = 0.5 \, {\rm nm}$ and $n_b = 0.1 \, {\rm M}$.
The $z=0$ surface is taken as the outer boundary of the proximal layer ($z=d$ in Fig.~\ref{fig1}).
}}
\label{figure2}
\end{figure*}

\subsection{Green's Function}
The Green's function defined in Eq.~(\ref{l4}) for the action $S$ of Eq.~(\ref{m10a}) satisfies
\begin{eqnarray}
\label{g1}
\left[ -\frac{\beta}{4\pi}\nabla\cdot\Big( \varepsilon(z)\nabla \Big) + \gamma(z) \right] G\left(\vecr,\vecr'\right) = \delta\left(\vecr-\vecr'\right) \, ,
\end{eqnarray}
where
\begin{eqnarray}
\label{g2}
\gamma(z) &=& 2\lambda_0\beta^2e^2\cosh(\beta e\psi_{0}) \\
\nonumber& &+ \beta^2e^2\lambda_0\delta(z)\left( \chi_+\e^{-\beta e\psi_{0}} + \chi_-\e^{\beta e\psi_{0}}  \right) \, ,
\end{eqnarray}
and $\lambda_0$ is the zero-loop fugacity. 
The system is translational invariant in the transverse $(x,y)$ directions,
and we can use the Fourier-Bessel transform by integrating out the angular dependence in polar coordinates,
\begin{eqnarray}
\label{g3}
\nonumber G(\vecr,\vecr') &=& \frac{1}{4\pi^2}  \int \D{\bf k}  ~g(k ; z,z') \e^{i{\bf k} \cdot (\bm{\rho} - \bm{\rho}')} = \\
& & = \frac{1}{2\pi}\int dk\, k \,  g(k ; z,z') J_0(k \vert \bm{\rho} - \bm{\rho}'\vert) \, ,
\end{eqnarray}
where $J_0$ is the zeroth-order Bessel function of the 1st kind,
$\bm\rho = (x,y)$ is the in-plane radial vector and $k=|{\bf k}|$.

As noted after Eq.~(\ref{ep9}), similar adhesivities corresponds to weak MF potentials, $\beta e\psi_{0} \ll 1$,
and the solution for the Green's function to first order in $\psi_0$ (see Appendix B for details) is:
\begin{eqnarray}
\label{g6}
\nonumber & &g(k; z,z') =  \frac{2\pi}{\beta\varepsilon_w p}\left[ 1 + \xi(k) \right]\e^{kz-pz'}  \qquad\qquad~~~~~ z < 0 \, ,\\
\nonumber & &g(k; z,z') =  \frac{2\pi}{\beta\varepsilon_w p}\left[ \e^{-p|z-z'|} + \xi(k) \e^{-p(z+z')}\right]  ~~~~ z \geq 0 \, ,\\
\end{eqnarray}
where we have defined (see also Appendix B),
\begin{eqnarray}
\label{g8}
& &\xi(k) \equiv \frac{ \varepsilon_w p - \varepsilon_a k - \gamma_s}{\varepsilon_w p + \varepsilon_a k + \gamma_s} \, , \nonumber \\
\label{g8a}
& &\gamma_s \equiv \half\varepsilon_w\kd^2 \left[ \chi_+\left( 1 - \beta e\psi_s  \right) + \chi_-\left( 1 + \beta e\psi_s  \right) \right] \, ,
\end{eqnarray}
and recall that $p=\sqrt{k^2+\kd^2}$.
Because $\psi \sim \chi_+ - \chi_-$, we can write $\gamma_s$ to order ${\cal O} (\chi_+ - \chi_-)$ as
\begin{equation}
\gamma_s \simeq \half\varepsilon_w\kd^2 \left( \chi_+ + \chi_- \right)\,.
\end{equation}

Of special interest is the equal-point Green's function,
\begin{eqnarray}
\label{g9}
G(z) \equiv G(\vecr,\vecr) = \int_0^\Lambda\D k \frac{k}{\beta\varepsilon_wp}\left[ 1+\xi(k)\e^{-2pz} \right] \, ,
\end{eqnarray}
where $\Lambda = 2\sqrt\pi/a \sim a^{-1}$ is a microscopic cutoff and $a$ is the minimal distance of approach between ions, defined earlier.
Note that for $a \to 0$, the Green's function diverges.
This is an artifact of the electrostatic interaction $\sim 1/r$ between point-like ions, which diverges as $r\to 0$.
In real systems, the ions have finite size that introduces a minimal distance of approach.
However, this distance is usually much smaller than any other system length-scales and can be taken safely to zero in many cases.

\subsection{Fugacities}
The ionic profiles can be calculated from Eq.~(\ref{m10}), while the fugacity, $\lambda$,
is related to the bulk density, $n_b$, through Eq.~(\ref{m12}).
It is, therefore, necessary to compute the thermal average, $\left< \e^{\mp i\beta e\phi } \right>$.

Employing the expansion of Eq.~(\ref{l14}) and using once again the Hubbard-Stratonovich transformation, yields
\begin{eqnarray}
\label{id2}
\left< \e^{\mp i\beta e\phi}\right> &=& \exp\left[ \mp\beta e\psi - \frac{\beta^2e^2\ell}{2} \, G(z)\right] \\ \nonumber\\
\nonumber&=& \exp\left[ \mp\beta e\left(\psi_{0} +\ell\psi_1\right) - \frac{\beta^2e^2\ell}{2} \, G(z)\right]  \, .
\end{eqnarray}
We proceed to determine the bulk fugacity from Eqs.~(\ref{m12}), (\ref{m10}) and the above equation.
The bulk fugacity is determined by the constraint
that the densities at $z\to\infty$ should match the bulk density, $n_b$,
\begin{eqnarray}
\label{id3}
\nonumber n_b &=& n_{\pm}\left(z\to\infty\right) \simeq \lambda \e^{- \half\beta^2e^2\ell \, G(z\to\infty)} \\
&\simeq& \lambda_0\left( 1 + \ell \left[ \frac{\lambda_1}{\lambda_0} -  \half\beta^2e^2 \, G(z\to\infty)\right] \right)\, .
\end{eqnarray}
The above equation and Eq.~(\ref{g9}) for the Green's function, $G(z)$, give the zeroth-order contribution and the one-loop correction to the bulk fugacity,
\begin{eqnarray}
\label{id3aa}
& &\lambda_0 = \lambda_0^{(s)} = n_b \, ,\\
\nonumber& &\frac{\lambda_1}{\lambda_0} =  \half\beta^2e^2  \, G(z\to\infty) = \frac{\lb}{2}\int_0^{\Lambda} \D k \frac{k}{p} = \frac{\lb}{2} (\Lambda-\kd) \, ,
\end{eqnarray}
where $\beta e^2G(z\to\infty)/2 = e^2u_b/2$ is exactly half of the electrostatic energy needed for adding an ion to the bulk electrolyte solution.

A similar reasoning relates the one-loop surface fugacity to
the electrostatic energy, $e^2u_s$, required to place an ion onto the air/water interface, where salt is absent ($\kd=0$).
The one-loop correction to the surface fugacity is then easily obtained from Eq.~(\ref{g9}) as,
\begin{eqnarray}
\label{id3aaa}
\frac{\lambda_1^{(s)}}{\lambda_0} =  \half\beta^2e^2  \, G\left(z=0;\kd=0\right) = \frac{\varepsilon_w\lb\Lambda}{\varepsilon_w+\varepsilon_a} \, ,
\end{eqnarray}
where $\beta e^2G\left(z{=}0;\kd{=}0\right)/2 = e^2u_s/2$.

We note that the difference between $\lambda_1$ and $\lambda_1^{(s)}$ is an artifact
arising from the decomposition of the free-energy to bulk and surface terms.
The surface ions are treated as they are half in the water and half in the air
(see Ref.~\cite{levin} for the calculation of the ion self-energy at the surface).

At this stage, we can write the one-loop potential by
substituting the MF potential of Eq.~(\ref{ep11}), the Fourier-Bessel transform of the Green's function, Eqs.~(\ref{g3}) and (\ref{g6}),
as well as the above fugacity expression, Eqs.~(\ref{id3aa})-(\ref{id3aaa}), into Eq.~(\ref{ep11aa}). This then yields
%
\begin{eqnarray}
\label{ep12a}
\psi_1(z) &=& \frac{1}{4}\kd\lb\psi_{0}(z) \int\D k \frac{k}{p}  \\
\nonumber&\times& \Bigg[ \xi(k) \Bigg( \frac{1 - \e^{-2pz}}{2p} + \frac{\xi(0) + \e^{-2pz}}{2(p+\kd)} \Bigg) \\
\nonumber &-& \frac{1+\xi(0)}{\kd} \left( 1 + \xi(k) -  \frac{2\varepsilon_wp}{(\varepsilon_w+\varepsilon_a)k} \right)  \Bigg]\, .
\end{eqnarray}
The above expression is one of our important results and will be used in Sec.~III to obtain the ionic profiles.

\section{Results}
We first derive the expressions for the ionic profiles close to the surface
and the total amount of ions contained within the proximal region.
Then, we use the Gibbs adsorption isotherm to obtain the interfacial tension and compare it to previous results.

\begin{figure}
\centering
\includegraphics[scale=0.8]{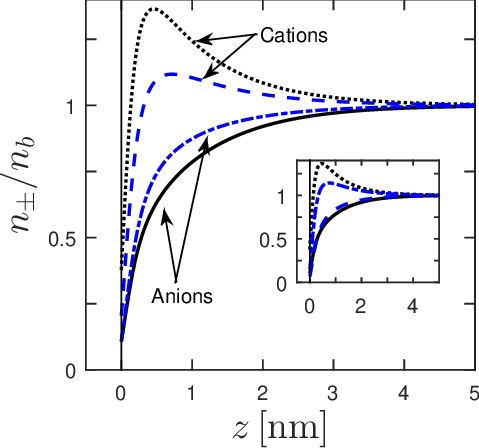}
\caption{ \textsf{ (color online).
One-loop ionic profiles for cations and anions obtained from Eq.~(\ref{id3b})
as function of the distance $z$ from the air/water interface for different values of $\alpha_{\pm}$.
The black lines are obtained for $\alpha_{+} = 0.1 \, \kbt$ and $\alpha_{-} = -1.0 \, \kbt$ (dotted for cations and solid for anions),
while the blue lines correspond to $\alpha_{+} = 0.1 \, \kbt$ and $\alpha_{-} = -0.5 \, \kbt$ (dashed for cations and dash-dotted for anions).
In the inset (same units), we present the profiles for $\alpha_{+} = 0.1 \, \kbt$, $\alpha_{-} = -1.0 \, \kbt$ (black dotted for cations and black solid for anions)
and $\alpha_{+} = 1.0 \, \kbt$, $\alpha_{-} = -0.1 \, \kbt$ (blue dash-dotted for cations and blue dashed for anions)
to show the asymmetry between positive and negative values of $\alpha_{+}$ and $\alpha_{-}$.
Other parameters are as in Fig.~\ref{figure2}.
The $z=0$ surface is taken as the outer boundary of the proximal layer
($z=d$ of Fig.~1).
}}
\label{figure3}
\end{figure}

\subsection{Ionic Profiles}
In order to obtain the analytical expression for the ionic profiles, we substitute the electrostatic potential, Eqs.~(\ref{ep11}) and (\ref{ep12a}),
together with the equal-point Green's function, Eq.~(\ref{g9}), into Eq.~(\ref{m10}).
The ion densities in water ($z>0$) as a function of the physical quantities, $n_b$ and $\chi_{\pm}$, are:
\begin{eqnarray}
\label{id3b}
\nonumber & &n_{\pm}(z) \simeq n_b \, \exp\Big[  - \half\beta^2e^2\ell \, G_s(z)\mp\beta e\left[\psi_{0}(z) +\ell\psi_1(z)\right]\Big] \\
& &\simeq n_b \, \e^{ - \half\beta^2e^2\ell \, G_s(z)} \Big( 1 \mp\beta e\left[\psi_{0}(z) +\ell\psi_1(z)\right]   \Big)   \, ,
\end{eqnarray}
where $n_{\pm}(z<0) = 0$ and $G_s(z) \equiv G(z) - G(z\to \infty)$.
We recall that the parameter $\ell$ should be set to unity at the end (see Sec.~II.B).
As we do not expand the densities to order ${\cal O} (\ell)$ (see Appendix C for further details), we need for consistency to re-exponentiate the fugacity expression,
$\lambda \simeq \lambda_0 + \ell\lambda_1 \simeq \lambda_0\exp\left(\ell\lambda_1/\lambda_0\right)$,
and use Eq.~(\ref{id3aa}).
In the second line of Eq.~(\ref{id3b}) we have expanded the exponent in order to keep only terms to order ${\cal O} (\chi_+ - \chi_-)$.
We recall that the model does not apply to the densities inside the proximal layer, $z \, \epsilon \, [0,d]$ of Fig.~\ref{fig1},
and the above equation is valid only outside this layer.

Although we cannot calculate the ionic profiles inside the proximal layer,
we can approximate the total number of cations/anions (per unit area) in this layer, defined as $N_{p}^{\pm}/A = \int_0^d \D z\, n_\pm(z)$.
In the spirit of Sec.~II.B, we expand Eq.~(\ref{m10a1}) to first order in $d$. Then, Eq.~(\ref{id3b}) is evaluated at $z=0^+$
and
used\footnote{In order to calculate macroscopic properties such as surface tension, one should use the {\it effective} densities $n^{\pm}_{s}$.
Similarly to Eq.~(\ref{id3b}), these effective densities can be written as,
%
\begin{eqnarray}
\label{id3ba}
\nonumber n_{\pm}^{(s)} &=& n_b\chi_{\pm} \, \exp\Big[ - \half\beta^2e^2\ell \, \tilde{G}_s(0) \mp\beta e\left[\psi_{0}(0) +\ell\psi_1(0)\right] \Big] \\
\nonumber&\simeq& n_b\chi_{\pm} \, \e^{ - \half\beta^2e^2\ell \, \tilde{G}_s(0)} \Big( 1 \mp\beta e\left[\psi_{0}(0) +\ell\psi_1(0)\right]   \Big)   \, .
\end{eqnarray}
where $\tilde{G}_s(0) \equiv G(z=0) - G\left(z=0;\kd=0\right)$.}
to derive $N_p^\pm$:
\begin{eqnarray}
\label{id3baa}
\nonumber\frac{N^{\pm}_{p}}{A} &\simeq& n_b d \exp \Big[-\beta\alpha_{\pm} - \half\beta^2e^2\ell \, {G}_s(0) \\
\nonumber& & \qquad\qquad\qquad\qquad\mp\beta e\left[\psi_{0}(0) +\ell\psi_1(0)\right] \Big]  \\
\nonumber&\simeq& n_b d \exp\Big[-\beta\alpha_{\pm} - \half\beta^2e^2\ell \, {G}_s(0)\Big] \\
& &\qquad\qquad\times\Big( 1 \mp\beta e\left[\psi_{0}(0) +\ell\psi_1(0)\right]   \Big)     \, ,
\end{eqnarray}
where in the last equality the electrostatic potential is expanded to first order since we are in the DH regime.

As explained after Eq.~(\ref{m10}), we treat the proximal layer as a collapsed surface layer lying at the water/air interface, $z=0$.
Namely, all distances are measured from the outer boundary of the proximal region, $z=d$, in the original system depicted in Fig.~\ref{fig1}.

\begin{figure*}[t]
\centering
\mbox{\hspace{-1cm}
   \subfigure{
   \includegraphics[scale = 0.7] {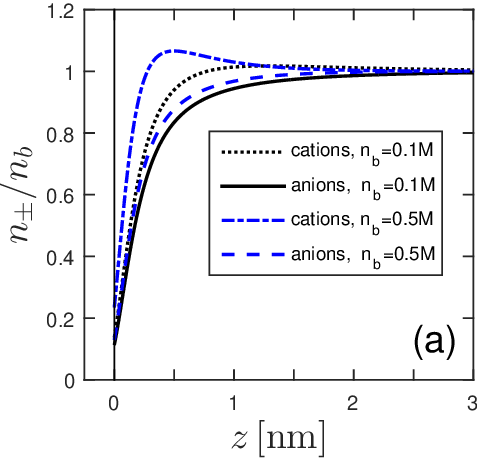}
   \label{fig4a}
 }\quad
   \subfigure{
   \includegraphics[scale = 0.7] {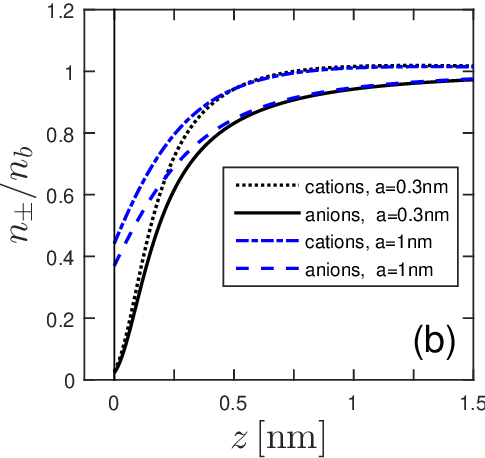}
   \label{fig4b}
 }\quad
 \subfigure{
   \includegraphics[scale = 0.7] {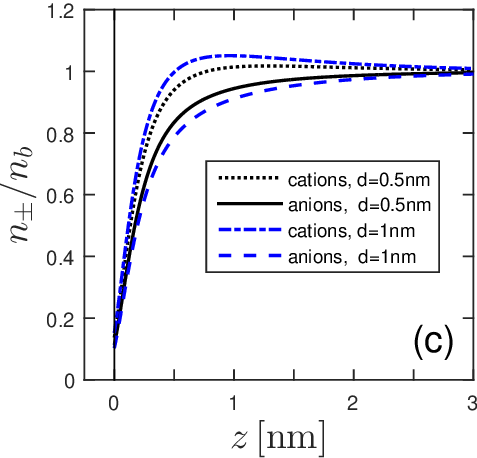}
   \label{fig4c}
 }}
\caption{\textsf{ (color online).
One-loop ionic profiles for cations and anions obtained from Eq.~(\ref{id3b})
as function of the distance $z$ from the air/water interface for: (a) different salt concentration, $n_b$, with $d = a = 0.5\, {\rm nm}$,
(b) different cutoff, $\Lambda = 2\sqrt{\pi}/a$, with $d = 0.5\, {\rm nm}$ and $n_b = 0.1 \, {\rm M}$,
and (c) different proximal layer width, $d$, with $a = 0.3 \, {\rm nm}$ and $n_b = 0.1 \, {\rm M}$.
Other parameters used are: $\alpha_\pm = \pm 0.1 \, \kbt$, and the rest are as in Fig.~\ref{figure2}.
The $z=0$ surface is taken as the outer boundary of the proximal layer ($z=d$ of Fig.~1).
}}
\label{figure4a}
\end{figure*}

In Fig.~\ref{figure2} we compare the obtained one-loop and MF ionic densities.
The one-loop correction is significant close to the dielectric discontinuity.
For $\alpha_\pm = \pm 0.1 \, \kbt$, the MF and one-loop concentrations coincide at $z \gtrsim 3d = 1.5 \, {\rm nm}$,
while for $\alpha_{+} = 0.1 \, \kbt$ and $\alpha_{-}=-0.5 \, \kbt$, the two profiles coincide at larger distances of $z \gtrsim 6d = 3 \, {\rm nm}$.
At these distances and above them, the calculated densities almost reach their bulk values.

The adhesivities in Fig.~\ref{figure2} correspond to repulsion of cations and attraction of anions from/to the proximal layer,
reflecting an effective negative surface charge density.
Therefore, the cations are attracted to an adjacent {\it ``secondary layer"} (sub-proximal) where they accumulate.
When the bias, $\beta|\alpha_{+} - \alpha_{-}|$, becomes larger,
the deviation of the one-loop profile from MF is noticeable even farther away from the surface.
However, the cations density peak, which corresponds to their accumulation at the secondary layer,
moves closer to the surface.
At distances larger than the peak position, $z>z^*$, the MF and one-loop densities differ only quantitatively,
whereas  the difference between the two is qualitative in the peak region.

Figure~\ref{figure3} presents different values of the adhesivities, $\alpha_{+} \neq \alpha_{-}$.
One can notice as a general trend that when the bias becomes larger,
the density peak of the
secondary layer increases in its height and shifts towards the air/water interface.
In the figure we show this trend by fixing $\alpha_{+} = 0.1 \, \kbt$
and plotting two different values for the anions adhesivity, $\alpha_{-} = -0.5 \, \kbt$ and $\alpha_{-} = -1.0 \, \kbt$.
For the former case, the density peak $n^* \simeq 1.36 n_b$ is at $z^* \simeq 0.45 \, {\rm nm}$,
while for the latter, the density peak $n^* \simeq 1.12 n_b$ is at $z^* \simeq 0.71 \, {\rm nm}$.

In Fig.~\ref{figure4a} we present the one-loop ionic profiles for (a) different salt concentration $n_b$,
(b) different cutoff $\Lambda=2\sqrt{\pi}/a$, and (c) different proximal layer width, $d$.
We use $\alpha_\pm = \pm 0.1 \, \kbt$, which yields an {\it effective} negative surface charge density.
For higher salt concentration, the ``second layer" (sub-proximal) peak of the cation density becomes
more pronounced and moves closer to the interface (as occurs for higher bias).
For concentration of $0.1$M, the density peak, $n^* \simeq 1.07 n_b$, is at $z^* \simeq  0.5 \, {\rm nm}$,
while for $0.5$M, it is located at $z^* \simeq  1.28 \, {\rm nm}$  with value of $n^* \simeq 1.02 n_b$.
The variation of anion concentration close to the surface increases with the salt concentration,
and both cation and anion profiles reach their bulk values closer to the surface.

As seen in Fig.~\ref{figure4a} (b), the different values of the cutoff, $\Lambda$,
only affect the Green's function as both $\psi_{0}$ and $\psi_1$ do not depend on the cutoff, $\Lambda$.
Therefore, different cutoffs only change the density very close to the interface,
and already at distances $z \gtrsim d = 0.5 \, {\rm nm}$, there is no difference between the ionic profiles for various cutoffs.
This means that our theory is quite robust for calculating ionic profiles and does not depend strongly on the value of the $\Lambda$ cutoff.

Figure~\ref{figure4a} (c) shows that by increasing the width of the proximal layer, $d$,
the density peak moves closer to the interface and the difference between the anions and cations profiles increases.
A similar trend is observed by increasing the salt concentration (see Fig.~\ref{figure4a} (a)),
but the dependence on $d$ is found to be somewhat weaker.
For proximal layer width of $d=0.5 \, {\rm nm}$, the density peak is at $z^* \simeq 0.94 \, {\rm nm}$  and its value is $n^* \simeq 1.05 n_b$,
while for $d = 1 \, {\rm nm}$, the peak is located at $z^* \simeq  1.25 \, {\rm nm}$  with value of $n^* \simeq 1.02 n_b$.
Furthermore, unlike the salt concentration effect, changing the proximal layer width reduces the profiles slope.
The limiting bulk values are obtained farther away from the surface and its proximal layer.


\subsection{Surface Tension}
%
We calculate the excess surface tension, $\Delta\gamma$, to one-loop order through the Gibbs adsorption isotherm
by using the one-loop ionic specific profiles for anions and cations.
In order to test our model, the results are compared with our previous work~\cite{EPL},
where the surface tension was calculated in a different way, directly from the free energy.
The latter method is thermodynamically equivalent to the Gibbs adsorption isotherm.
Note that $\alpha$ of Ref.~\cite{EPL} corresponds to $\alpha_{-}$ of the present paper, and in order to make the comparison we should set $\alpha_{+} = 0$.

The Gibbs adsorption isotherm gives the excess surface tension (with respect to the bare air/water interface),
\begin{eqnarray}
\label{st1}
\Delta\gamma = -\kbt \sum_{\pm} \int_0^{n_b^{(\pm)}} \frac{\D s}{s} \int\D z\left[ n_{\pm}(z,s) - s \right] \, .
\end{eqnarray}
For $n_{\pm}(z)$ we use Eq.~(\ref{id3b}) and the effective surface densities, $n_{s}^{\pm}$ (see footnote before Eq.~(\ref{id3ba})),
where the integration is performed on the bulk ionic concentration, $n_b$.

Although the loop expansion is not fully justified for the densities, it is valid for free energy
and other macroscopic quantities such as the surface tension, even in presence of a dielectric discontinuity
(see Appendix C).
When calculating free energies (or surface tension), one has to expand all terms to first order in $\ell$.
Hence, we expand the surface tension to first order in $\ell$
and write the one-loop surface tension in the DH regime, ${\cal O} \left( \chi_+ - \chi_- \right)$, as
\begin{eqnarray}
\label{st3}
\nonumber\Delta\gamma &\simeq& \Delta\gamma_0 + \ell\Delta\gamma_1 = - n_b\left( \chi_+ + \chi_- \right) \\
\nonumber&+& \ell \int_0^{n_b}\!\!\D n^{\prime} \, \frac{\lb}{2} \int_0^{\Lambda}\D k
\nonumber\Bigg( \frac{k\xi\left(k,n^{\prime}\right)}{p^2(n^{\prime})}  \\
&+&  \frac{k}{p}\Big[ 1 + \xi\left(k,n^{\prime}\right)\Big] - \frac{2\varepsilon_w}{\varepsilon_w+\varepsilon_a} \Bigg) \, ,
\end{eqnarray}
where $\ell$ is set to unity.

\begin{figure}[t]
\centering
\includegraphics[scale=0.8]{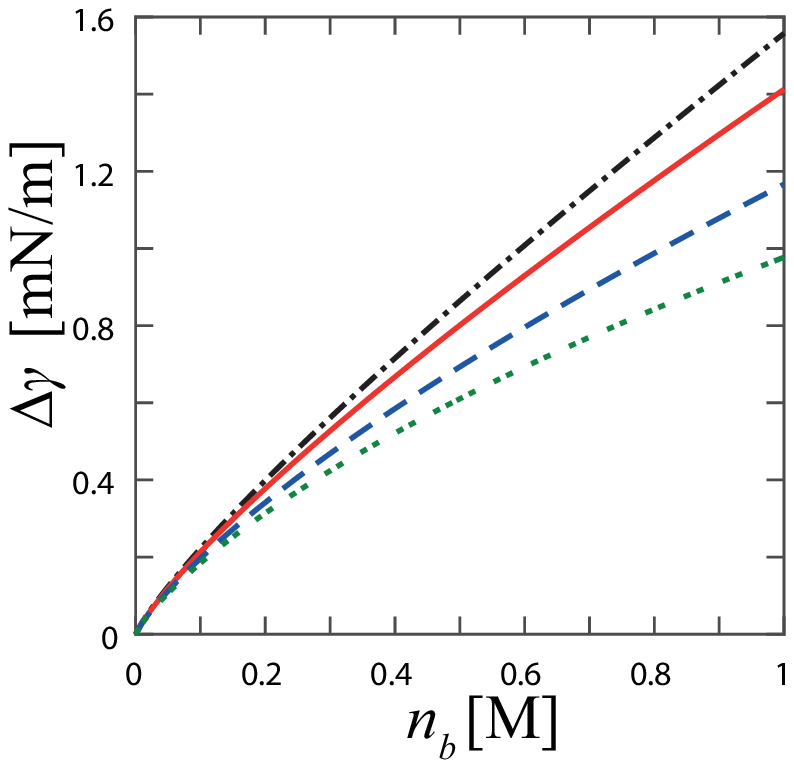}
\caption{ \textsf{ Surface tension as a function of the salt bulk concentration,
as calculated from the Gibbs adsorption isotherm of Eq.~(\ref{st3}).
These lines are in perfect agreement with the surface tension calculated directly from the free energy, Ref.~\cite{EPL}.
The different lines refer to different values of  $\alpha_{-}\equiv\alpha$ and $d$ (while keeping $\alpha_{+}=0$).
The comparison is done for the same parameters are used as in Ref.~\cite{EPL}:
$\alpha = 0.18 \, \kbt$ and $d=0.71 \, {\rm nm}$ (black dash-dotted line) as was fitted in Ref.~\cite{EPL} for NaF,
$\alpha = 0.135 \, \kbt$ and $d=0.69 \, {\rm nm}$ (red solid line) for NaCl,
$\alpha = 0.069 \, \kbt$ and $d=0.69 \, {\rm nm}$ (blue dashed line) for NaBr,
and  $\alpha = 0.023 \, \kbt$ $d=0.69 \, {\rm nm}$ (green double-dotted line) for NaI.
   } }
\label{figure5}
\end{figure}

Our results are shown in Fig.~\ref{figure5}, and are indistinguishable from those of Ref.~\cite{EPL}.
In fact, we have compared numerically the surface tension of Eq.~(\ref{st3}) with Eq.~(24) of Ref.~\cite{EPL},
and they are equal for the same values of $\alpha_{\pm}$ (or equivalently, $\chi_{\pm}$).

We would like to emphasize that the surface tension results can be used to predict quantitatively the corresponding ionic densities.
The surface tension can be measured with commonly available techniques (such as the drop volume technique).
Then, the experimental results can be fitted with Eq.~(\ref{st3}) (or Eq.~(24) of Ref.~\cite{EPL}, which are the same for $\alpha_{+}=0$)
in order to obtain the numerical values of the adhesivity parameter, $\alpha_{\pm}$.
The ionic densities can then be obtained from Eqs.~(\ref{id3b}) and (\ref{id3ba}), where the only fitting parameter is the adhesivity, $\alpha_\pm$.

\section{Discussion}
Figure~\ref{figure2} shows large difference between our obtained one-loop ionic profiles and the corresponding MF profiles.
This difference is a result of the image-charge interaction~\cite{onsager_samaras},
which gives rise to strong repulsion of ions from the surface at short distances, as demonstrated by the Green's function
(see $G_s(z)$ of Eqs.~(\ref{id3b}) and (\ref{id3baa})).
This repulsion depends on the adhesivity, and can be magnified ($\alpha_{\pm} > 0$) or reduced ($\alpha_{\pm} < 0$). This is due to correlation effects that couple
the ionic adhesivities with their image-charge interaction at the surface.
Although at very small distances our continuum theory is not accurate,
it gives a qualitatively correct behavior, which is not the case for the MF profiles.

The one-loop correction has a strong contribution close to the interface.
This is due to the Green's function that largely affects the density profile close to the interface.
Because the Green's function is the same for cations and anions,
the cation/anion profiles at short distances (before the apparent density peak) are rather similar.
For larger distances, the Green's function decays exponentially,
and the deviation becomes larger as function of the $\alpha_{+} \neq \alpha_{-}$ bias.

On the other hand, the one-loop correction to the electrostatic potential has an opposite effect on cations and anions.
Thus, it results in an increased deviation from the cation/anion MF profiles at intermediate distances from the interface.
For a small bias, $\alpha_{+} \simeq \alpha_{-}$, the density profiles calculated from one-loop and MF coincide before reaching their bulk values,
while for large bias, $\alpha_{+} \neq \alpha_{-}$, they only coincide farther from the surface, when they reach their equal bulk value, $n_b$.
This effect is mainly due to the one-loop correction of the electrostatic potential.
Such a correction strongly depends on the bias,
and has a longer-range effect than the Green's function.
%
The electrostatic potential (MF and one-loop) also has an exponential decay, but slower than that of the Green's function.
The range of the one-loop correction increases with the bias,
leading to quantitative deviation from the MF profile.

Another remark is that there is no symmetry between positive (repulsive) and negative (attractive) values of $\alpha_{\pm}$.
The dependence of $\chi_{\pm}$  on the adhesivity through $\exp(-\beta\alpha_{\pm})$
is asymmetric with respect to $\alpha_{\pm} \leftrightarrow -\alpha_\mp$.
This is clearly seen in the inset of Fig.~\ref{figure3},
where we compare the profiles for the same bias
$\alpha_{+} = 0.1 \, \kbt$, $\alpha_{-} = -1.0 \, \kbt$ with the opposite case of $\alpha_{+} = 1.0 \, \kbt$, $\alpha_{-} = -0.1 \, \kbt$.
The two cases show quite different profiles.
The density peak of ions accumulated in the ``secondary layer" is higher and closer to the air/water interface for the first case.
This observation implies that negative values of $\alpha_\pm$ have stronger effect than the positive ones.

For repulsive and large ion-surface interaction, $\alpha_{\pm} > 5 \, \kbt$, there is no observable difference
between different ionic profiles
because the ions are completely expelled from the proximal layer.
This clearly is qualitatively very different than the case of attractive ion-surface interaction. In the latter case,
the number of ions in the proximal layer increases as $\alpha_\pm$ becomes more negative.
The number of ions at the surface is not bounded as we assume the ion-surface is repulsive or slightly attractive.
However, for highly negative $\alpha_{\pm}$, our theory is violated as the weak coupling limit we employed is not valid anymore.
Furthermore, for such strong adsorption the steric repulsion at the proximal layer has to be included, and will give a bound to the number of ions at the surface.

\subsection{Comparison with Previous Models}
\begin{figure}[t]
\centering
\includegraphics[scale=0.8]{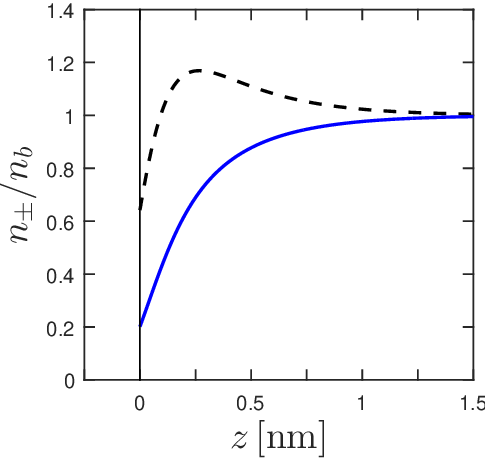}
\caption{ \textsf{ (color online).
Ionic profiles for NaI obtained from Eq.~(\ref{id3b}) (one-loop) with $\alpha_{\pm}$ taken from the fitting parameters in Table II of Ref.~\cite{JCP}:
$\alpha_{+} = \alpha_{_{\rm Na}} =  0.11 \, \kbt$, $\alpha_{-} = \alpha_{_{\rm I}} = -0.071\, \kbt$.
The black dashed line and the solid blue line correspond to the ${\rm Na}^+$ and ${\rm I}^-$ ions, respectively.
Other parameters used are: $d = a = 0.69 \, {\rm nm}$, $n_b = 1 \, {\rm M}$ and the rest are as in Fig.~\ref{figure2}. The profiles compare quite favorably with those reported in Refs.~\cite{Netz2010} and \cite{levin2}.
}}
\label{figure6}
\end{figure}

In Fig.~\ref{figure6} we compare our predictions for the ionic profiles with those obtained by Netz and co-workers (Fig.~3 B of Ref.~\cite{Netz2010}),
and those of Levin and co-workers (Fig.~3 of Ref.~\cite{levin2}).
In general, our analytical results compare favorably with both previous models.

Figure~\ref{figure6} shows our computed ionic profiles, Eq.~(\ref{id3b}), for NaI as obtained with the fitting parameters of Table II, Ref.~\cite{JCP}:
$\alpha_{+} = \alpha_{_{\rm Na}} =  0.11 \, \kbt$, $\alpha_{-} = \alpha_{_{\rm I}} = -0.071 \, \kbt$.
The density peak of Fig.~\ref{figure6}, related to the accumulation of ${\rm Na}^+$ ions
(dashed line) in the secondary layer, lies at $z^*\simeq 0.25 \, {\rm nm}$,
and its value is $n^* \simeq 1.2 n_b$.
The density profiles reach their bulk value at $z \simeq 1.5 \, {\rm nm}$.
As explained after Eq.~(\ref{id3baa}), we treat the proximal layer as a collapsed surface layer lying at $z=0$.
Namely, all distances are measured from the outer boundary of the proximal layer.

The results in Fig.~\ref{figure6} are in good agreement with those of Ref.~\cite{Netz2010}.
As our proximal layer is collapsed onto $z=0$,
only distances  away from the first density peak of ${\rm I}^-$ in Fig.~3~B of Ref.~\cite{Netz2010} should be compared to ours.
The ${\rm Na}^+$ density peak of Ref.~\cite{Netz2010} is at  distance $z\simeq 0.3 \, {\rm nm}$ away from the ${\rm I}^-$ peak,
and the ionic profiles reach their bulk values at distance $z\simeq 1.4 \, {\rm nm}$ away from the ${\rm I}^-$ peak.
Moreover, the height of the ${\rm Na}^+$ density peak is $n^* \simeq 1.2 n_b$. All these findings agree with our results (Fig.~\ref{figure6}).
The ${\rm I}^-$ profile becomes similar to ours only at distances $ z\gtrsim 0.4 \, {\rm nm} \simeq 0.58 \cdot d $ away from the proximal layer. This
is not surprising as our results are not very accurate for small distances, $z \lesssim d$.

Our profiles looks also quite similar to those of Ref.~\cite{levin2}. Again, one should compare distances only outside the excluded region
(where the ${\rm Na}^+$ density vanishes in Fig.~3 of Ref.~\cite{levin2}), and not within the proximal layer of the interface (the Gibbs dividing surface).
The ${\rm Na}^+$ density peak of Ref.~\cite{levin2} is located at $z^* \simeq 0.35 \, {\rm nm}$ away from the excluded region,
has a peak value of $n^* \simeq 1.1 n_b$, while the ionic profiles reach their bulk values at $z\simeq 1.5 \, {\rm nm}$.
%
%
%

\section{Conclusions}
We have presented a model for ionic profiles in the proximity of an interface that has a sharp dielectric jump.
We considered separately ionic-specific interactions for anions and cations, modeled by two adhesivity parameters, $\alpha_\pm$.
These added surface interactions are formulated as a self-consistent non-linear boundary condition.

Ionic densities are calculated analytically from one-loop order of the free energy in the Debye-H\"uckel (DH) regime,
assuming that the surface-induced bias towards one of the ionic species, $\beta|\alpha_{-} - \alpha_{+}|$, is small.
The theory is less applicable for high biases because of the limitation of the linear DH regime.
However, one could apply the same formalism
to the full non-linear PB with the same non-linear boundary condition and the appropriate Green's function~\cite{markovich_acids}. It is more cumbersome but doable.

In order to simplify the calculation, we require that the proximal layer is microscopically  small.
For proximal layer with width comparable with the minimal distance of approach of the ions, $d \simeq a$,
there is not much sense in discussing the profiles inside the proximal layer. It is straightforward but tedious to calculate analytically the ionic profiles for proximal layers of finite thickness. This will allow calculations of the ionic profiles inside the proximal layer.

It is important to stress that we do not use the regular loop-expansion for the densities, as it fails close to a dielectric jump.
Instead, we expand the free energy to one-loop order and compute the corresponding ionic profiles from it.
This re-summation of the loop expansion is equivalent to a cumulant-expansion around a non-zero value of the electrostatic potential.

Our analytical results present ionic-specific profiles close to dielectric discontinuities,
and are in agreement with previous ionic profiles obtained from simulations~\cite{Netz2010}
and numerical calculations~\cite{levin2}.
We have recovered the same surface tension results as in Refs.~\cite{EPL} and \cite{JCP}, and they correspond to the reverse Hofmeister series.
The method gives precisely the same results as from direct free energy calculations of the surface tension~\cite{EPL,JCP}.
Therefore, one can use the fitted $\alpha_{\pm}$ parameters, determined by macroscopic surface-tension measurements of specific electrolyte solutions, to obtain their ionic profiles close to the air/water interface.
The same calculations can be also performed at the oil/water interface, with other fitted values of the adhesivity, $\alpha_\pm$.

It is possible to determine the adhesivities values, $\alpha_{\pm}$, from quantum chemistry simulations,
or from a more microscopic approach.
As explained above, in a coarse-grained theory
one averages over length-scales of the order of the ionic size,
which is also the characteristic length-scale of the ionic specific potential, $V_{\pm}$~\cite{levin2,JCP}.
%
In the future, it will be of merit to calculate from the ionic profiles,
other macroscopic quantities (beside the surface tension), such as
the differential capacitance~\cite{nakayama},
the solution dielectric constant~\cite{levy} and the solution viscosity~\cite{visc}.

\bigskip

{\it Acknowledgements.~}
We thank A. Cohen and R. Podgornik for many insightful discussions, and U. Sivan for raising questions that started this project.
This work was supported in part by the Israel
Science Foundation (ISF) under Grant No. 438/12,
the US-Israel Binational Science Foundation (BSF) under
Grant No. 2012/060, and the ISF-NSFC joint research program under Grant No.  885/15. DA would like to thank the hospitality of IPhT at CEA-Saclay, where this work has been completed and acknowledges financial support from the French CNRS.

\bigskip

\appendix


\section{Electrostatic Potential Derivation within One-Loop}
We elaborate here on the general formalism for calculating the electrostatic potential to one-loop order.
The action for Coulombic systems is,
\begin{eqnarray}
\label{ep1}
S\left[\phi\right] = \int\D\vecr \Bigg[ \frac{\beta\varepsilon({\bf r})}{8\pi}[\nabla \phi({\bf r})]^2 + f(\vecr;\phi,\lambda_i) \Bigg] \, ,
\end{eqnarray}
where $f(\vecr;\phi,\lambda_i)$ includes the entropy and the fixed charge terms,
with $\lambda_i$ being the fugacities of the $i^{\rm th}$ species.
By using  the one-loop expansion of $\Gamma$, Eq.~(\ref{l14}), we write the equation of state, Eq.~(\ref{l11}) as,
\begin{eqnarray}
\label{ep2}
\frac{\delta S(\psi)}{\delta\psi(\vecr)} + \frac{\ell}{2}{\rm Tr} \, G(\vecr,\vecr';\psi)\frac{\delta S_2(\vecr,\vecr';\psi)}{\delta\psi(\vecr)} = 0 \, .
\end{eqnarray}
The electrostatic potential is completely determined by this equation of state.
We proceed by expanding the fugacities and electrostatic potential, as explained after Eq.~(\ref{l4aa}),
\begin{eqnarray}
\label{ep3}
\nonumber & &\psi = \psi_{0} +\ell\psi_1 \, , \\
& &\lambda_i = \lambda^{(i)}_0 + \ell\lambda^{(i)}_1 \, .
\end{eqnarray}
Then, by substituting Eq.~(\ref{ep1}) into Eq.~(\ref{ep2}), two equations are obtained.
The first is for the saddle-point potential,
\begin{eqnarray}
\label{ep4}
\frac{\beta\varepsilon}{4\pi}\nabla^2\psi_{0} + \frac{\partial f(\psi,\lambda_i,\vecr)}{\partial\psi}\Big|_{\lambda^{(i)}_0,\psi_{0}} = 0 \, ,
\end{eqnarray}
and the second is for the one-loop correction,
\begin{eqnarray}
\label{ep5}
\frac{\beta\varepsilon}{4\pi}\nabla^2\psi_1 &+& \frac{\partial^2 f(\phi,\lambda_i,\vecr)}{\partial\psi^2}\Big|_{\lambda^{(i)}_0,\psi_{0}}\psi_1 \\
\nonumber&+& \sum_i \frac{\partial^2 f(\psi,\lambda_i,\vecr)}{\partial\psi\lambda_i}\Big|_{\lambda^{(i)}_0,\psi_{0}}\lambda^{(i)}_1 \\
\nonumber&-& \half G(\vecr,\vecr)\frac{\partial^3 f(\psi,\lambda_i,\vecr)}{\partial\psi^3}\Big|_{\lambda^{(i)}_0,\psi_{0}}  = 0 \, .
\end{eqnarray}

The saddle-point equation, Eq.~(\ref{ep4}), is obtained from the variation principle, $\delta S/\delta\psi = 0$, and gives a modified PB equation.
We use the relation between $S_2(\vecr,\vecr')$ and $G(\vecr,\vecr')$ from Eq.~(\ref{l4})
to write $\psi_1$, the one-loop correction of Eq.~(\ref{ep5}), in the form,
\begin{eqnarray}
\label{ep6}
\nonumber\psi_1(\vecr) &=& \int\D\vecr'G(\vecr,\vecr';\psi_{0})
\Bigg[ \sum_i \frac{\partial^2 f(\psi,\lambda_i,\vecr')}{\partial\psi\partial\lambda_i}\Big|_{\lambda^{(i)}_0,\psi_{0}}\lambda^{(i)}_1 \\
&-& \half G(\vecr',\vecr')\frac{\partial^3 f(\psi,\lambda_i,\vecr')}{\partial\psi^3}\Big|_{\lambda^{(i)}_0,\psi_{0}}  \Bigg] \, .
\end{eqnarray}

To connect these general results with the present study, we use Eqs.~(\ref{m10a}) and (\ref{ep1}) to get the form of $f(\psi,\lambda_i,\vecr)$,
\begin{eqnarray}
\label{ep7}
\nonumber f(\psi,\lambda,\vecr) &=& -2\lambda\cosh\left[\beta e\psi(\vecr)\right] \\
&-& \delta(z)\lambda_s\left( \chi_+ \e^{-\beta e\psi({\bf r})}
+ \chi_{-} \e^{\beta e\psi({\bf r})} \right) \, ,
\end{eqnarray}
where $\lambda=\lambda_\pm$ is the bulk fugacity and $\lambda_s$ is the surface one.
The zeroth order in the loop expansion, Eq.~(\ref{ep4}), with Eq.~(\ref{ep7}), gives the PB equation for planar geometry
\begin{eqnarray}
\label{a8}
\nonumber & &\psi_{0}^{\prime\prime}(z) = 0 \qquad\qquad\qquad\qquad\quad\,\,\,\, z<0 \, ,\\
& &\psi_{0}^{\prime\prime}(z) = \frac{8\pi e \lambda_0}{\varepsilon_w} \sinh\left(\beta e\psi_{0}\right) \qquad z > 0\, ,
\end{eqnarray}
with a special boundary condition at $z=0$,
\begin{eqnarray}
\label{a9}
\nonumber \varepsilon_w\psi_{0}^\prime|_{_{0^+}} - \varepsilon_a \psi_{0}^\prime|_{_{0^-}} = -4\pi e \lambda_0 \left( \chi_{+}\e^{-\beta e\psi_s} - \chi_{-}\e^{\beta e\psi_s} \right)  \, . \\
\end{eqnarray}

To obtain the one-loop potential, we substitute Eq.~(\ref{ep7}) into Eq.~(\ref{ep6}), yielding
%
\begin{eqnarray}
\label{a10}
\nonumber\psi_1(\vecr) &=& \beta e \int\D\vecr' \,  G(\vecr,\vecr') \Bigg[ 2\lambda_0\sinh[\beta e\psi_{0}(\vecr')] \\
\nonumber&\times& \left( \half\beta^2e^2G(\vecr',\vecr') - \frac{\lambda_1}{\lambda_0} \right) \\
\nonumber&-& \delta(z')\lambda_0\Big( \chi_+\e^{-\beta e\psi_{0}} -\chi_-\e^{\beta e\psi_{0}} \Big) \\
\nonumber &\times& \Big( \half\beta^2e^2G(\vecr',\vecr') - \frac{\lambda_1^{(s)}}{\lambda_0} \Big) \Bigg] \, . \\
\end{eqnarray}
%
Equations~(\ref{a8})-(\ref{a10}) are used in Sec.~II.C to calculate the electrostatic potential.

\section{Green's Function in the DH regime}

Since we are in the DH regime, using Eqs.~(\ref{g1})-(\ref{g3}) the equations for $g(k; z_1,z_2)$ to first order in $\psi_0$ are:
\begin{eqnarray}
\label{g4}
& &\nonumber g''(k; z_1,z_2) - k^2g(k; z_1,z_2) = 0 \qquad\qquad\qquad\quad z_1 < 0  \, ,\\
& &\nonumber g''(k; z_1,z_2) - p^2g(k; z_1,z_2) = -\frac{4\pi}{\beta\varepsilon_w}\delta(z_1-z_2)  \quad  z_1 \geq 0  \, , \\
\end{eqnarray}
where $g'=dg/dz$, $g''=d^2g/dz^2$ and $p^2 = k^2 + \kd^2$.
The boundary conditions for $g$ are:
\begin{eqnarray}
\label{g5}
& &\nonumber \varepsilon_w g'(k; 0^+,z_2) - \varepsilon_a g'(k; 0^-,z_2) = \gamma_s g(k; 0,z_2)  \, ,\\
& &\nonumber g'(k; z_1^{+},z_1) - g'(k; z_1^{-},z_1) = -\frac{4\pi}{\beta\varepsilon_w} \, ,\\
& &g'(k; z_1\to\pm\infty,z_2) = 0 \, ,
\end{eqnarray}
with $z_1 \geq 0$, and $\gamma_s$ is defined in Eq.~(\ref{g8a}).

The solution for the Green's function is then,
\begin{eqnarray}
\label{g6a}
\nonumber & &g(k; z_1,z_2) =  \frac{2\pi}{\beta\varepsilon_w p}\left[ 1 + \xi(k) \right]\e^{kz_1-pz_2}  \qquad\qquad~~~~~ z_1 < 0 \, ,\\
\nonumber & &g(k; z_1,z_2) =  \frac{2\pi}{\beta\varepsilon_w p}\left[ \e^{-p|z_1-z_2|} + \xi(k) \e^{-p(z_1+z_2)}\right]  ~~~~ z_1 \geq 0 \, , \\
\end{eqnarray}
where $\xi(k)$ defined in Eq.~(\ref{g8}) is repeated here for convenience
\begin{eqnarray}
& &\xi(k) \equiv \frac{ \varepsilon_w p - \varepsilon_a k - \gamma_s}{\varepsilon_w p + \varepsilon_a k + \gamma_s} \, , \nonumber \\
& &\gamma_s \equiv \half\varepsilon_w\kd^2 \left[ \chi_+\left( 1 - \beta e\psi_s  \right) + \chi_-\left( 1 + \beta e\psi_s  \right) \right] \, ,
\end{eqnarray}
We will use the Green's function solution, Eq.~(\ref{g6a}) to compute the one-loop correction to the electrostatic potential
and the fugacities (Sec.~II.E), and then to calculate the one-loop ionic densities (Sec.~III.A).


\section{Limitations of the One-Loop Expansion}

It is known that the loop expansion method has problems close to dielectric discontinuities~\cite{dean2004,sahin2012}.
The small parameter in such an expansion of the grand-partition function
depends on the system under consideration (see, {\it e.g.,} Ref.~\cite{Netz_Moriera}).
The expansion validity is determined by its coupling parameters~\cite{podgornik_review,markovich_SC},
and is not related to the existence of a dielectric jump.

The problem arises when one calculates microscopic quantities, such as ionic profiles, beyond the MF approximation (zeroth-loop order),
because there is {\it no guarantee} that this is a valid expansion for these quantities.
As we will show below, the ionic densities can become negative ({\it i.e.}, nonphysical) close to the interface, when expanded to one-loop order.
This is not the case when calculating macroscopic properties, such as the surface tension,
for which the loop-expansion validity is determined by the coupling parameters.

Our results for the density profiles, Eqs.~(\ref{id3b}) and (\ref{id3ba}), are exact to one-loop order in the free-energy.
Nevertheless, in a consistent loop-expansion one should expand the exponents in the above equations to first order in $\ell$.
For example, Eq.~(\ref{id3b}), yields
\begin{eqnarray}
\label{id4}
\nonumber n_{\pm}(z) &\simeq& n_b\e^{\mp\beta e\psi_{0}(z)} \left[ 1 +\ell\left( \mp\beta e\psi_1(z) - \frac{\beta^2e^2}{2} \, G_s(z) \right)\right] \, , \\
\end{eqnarray}
where $G_s(z)$ includes the one-loop correction to the fugacity, as shown in Eq.~(\ref{id3aa}).
The problem with this expansion is that $G_s$ diverges as $z\to0$,
$G_s(0) \sim \Lambda (\varepsilon_w - \varepsilon_a)/(\varepsilon_w + \varepsilon_a) \sim 1/a \to\infty$.
This gives negative (nonphysical) densities at small values of $z$,
and is a well-known deficiency of the loop-expansion~\cite{dean2004,sahin2012}.

The failure of the regular loop-expansion is an artifact of the sharp dielectric jump at the air/water interface.
For this sharp jump, we do not expand the densities to first order in $\ell$, which amounts to perform a cumulant-expansion around
a non-zero value of the electrostatic potential.




\begin{thebibliography}{99}

\bibitem{hofmeister}
W. Kunz, J. Henle, and B.W. Ninham, Curr. Opin. Coll.  Interface Sci. {\bf 9}, 19 (2004).

\bibitem{collins1985}
K. D. Collins and M. W. Washabaugh, Q. Rev. Biophys. {\bf 18}, 323 (1985).

\bibitem{ruckenstein2003a}
M. Manciu, and E. Ruckenstein, Adv. Colloid Interface Sci. {\bf 105}, 63 (2003).

\bibitem{kunz2010}
W. Kunz,  Curr. Opin. Coll. Interface Sci. {\bf 15}, 34 (2010).

\bibitem{Sivan2009}
M. Dishon, O. Zohar, and U. Sivan, Langmuir {\bf 25}, 2831 (2009).

\bibitem{Sivan2013}
J. Morag, M. Dishon, and U. Sivan, Langmuir, {\bf 29}, 6317 (2013).

\bibitem{pashely}
R. M. Pashley,  J. Coll. Interface Sci. {\bf 83}, 531 (1981).

\bibitem{parsegian1992}
D. C. Rau and V. A. Parsegian, Biophys. J. {\bf 61}, 260 (1992).

\bibitem{parsegian1994}
R. Podgornik, D. Rau, and V. A. Parsegian, Biophys. J. {\bf 66}, 962 (1994).

\bibitem{air_water_2}
F. A. Long and G. C. Nutting, J. Am. Chem. Soc. {\bf 64}, 2476 (1942).

\bibitem{air_water_3}
J. Ralston and T. W. Healy, J. Coll. Interface Sci. {\bf 42}, 1473 (1973).

\bibitem{Wagner}
C. Wagner, Phys. Z. {\bf 25}, 474 (1924).

\bibitem{onsager_samaras}
L. Onsager and N. N. T. Samaras, J. Chem. Phys. {\bf 2}, 628 (1934).

\bibitem {Debye1923} P.~W. Debye and E. H\"uckel, Phys. Z. {\bf 24}, 185 (1923).

\bibitem{Kunz_Book}
W. Kunz, {\it Specific Ion Effects} (World Scientific, Singapore, 2009).

\bibitem{Dan2011}
D. Ben-Yaakov, D. Andelman, R. Podgornik, and D. Harries,
Curr. Opin. Coll.  Interface Sci. {\bf 16}, 542 (2011).

\bibitem{EPL}
T. Markovich, D. Andelman, and R. Podgornik, Europhys. Lett. {\bf 106}, 16002 (2014).

\bibitem{JCP}
T. Markovich, D. Andelman, and R. Podgornik, J. Chem. Phys. {\bf 142}, 044702 (2015).

\bibitem{Netz2012} N. Schwierz and  R. R. Netz, Langmuir {\bf 28}, 3881 (2012).

\bibitem{Netz2013} N. Schwierz, D. Horinek and  R. R. Netz, Langmuir {\bf 29} 2602 (2013).

\bibitem{Netz2010} N. Schwierz, D. Horinek and  R. R. Netz, Langmuir {\bf 26}, 7370 (2010).

\bibitem{levin2}
Y. Levin, A. P. dos Santos, and A. Diehl, Phys. Rev. Lett. {\bf 103}, 257802 (2009).

\bibitem{dean2004}
D. S. Dean and R. R. Horgan, Phys. Rev. E {\bf 70}, 011101 (2004).

\bibitem{sahin2012}
S. Buyukdagli, C. V. Achim and T. Ala-Nissila, J. Chem. Phys. {\bf 137}, 104902 (2012).

\bibitem{netz2003}
R. R. Netz and H. Orland, Eur. Phys. J. E {\bf 11}, 301–311 (2003).

\bibitem{sahin2010}
S. Buyukdagli, M. Manghi and J. Palmeri, Phys. Rev. E {\bf 81}, 041601, (2010).

\bibitem{dean2004st}
D. S. Dean and R. R. Horgan, Phys. Rev. E {\bf 69}, 061603 (2004).

\bibitem{podgornik1988}
R. Podgornik and B. Zeks, J. Chem. Soc. {\bf 84}, 611 (1988).

\bibitem{netz2000}
R. R. Netz and H. Orland, Eur. Phys. J. E {\bf 1}, 203 (2000).

\bibitem{gravity}
I. L. Buchbinder, S. Odintsov and L. Shapiro, {\it Effective action in quantum gravity} (CRC Press, 1992).‏

\bibitem{levin}
Y. Levin, Phys. Rev. Lett. {\bf 102}, 147803 (2009).

\bibitem{markovich_acids}
T. Markovich, D. Andelman and R. Podgornik unpublished.

\bibitem{nakayama}
Y. Nakayama, and D. Andelman, J. Chem. Phys. {\bf 142}, 044706 (2015).

\bibitem{levy}
A. Levy, D. Andelman and H. Orland, Phys. Rev. Lett. {\bf 108}, 227801 (2012).

\bibitem{visc}
B. Hribar, N. T. Southall, V. Vlachy and K. A. Dill, J. Am. Chem. Soc. {\bf 124}, 12302 (2002).

\bibitem{Netz_Moriera}
A. G. Moriera and R. R. Netz, Europhys. Lett., {\bf 52}, 705 (2000).

\bibitem{podgornik_review}
A. Naji, M. Kanduc, J. Forsman and R. Podgornik, J. Chem. Phys. {\bf 139}, 150901 (2013).

\bibitem{markovich_SC}
T. Markovich, R. M. Adar and D. Andelman, to be published.


%
%
%


\end{thebibliography}
\end{document}